\documentclass[conference]{IEEEtran}
\usepackage{tikz}

\usepackage{cite}
\usepackage{amsmath,amssymb,amsfonts}
\usepackage{algorithmic}
\usepackage{graphicx}
\usepackage{textcomp}
\usepackage{xcolor}
\usepackage{algorithm}
\usepackage{tikz-network}
\usepackage{pgfplots,pgfplotstable}
\usepackage{comment}
\usetikzlibrary{shapes.geometric, arrows}

\usepackage{booktabs}
\usepackage{caption}
\usepackage{multicol}

\usepackage{subcaption}  
\usepackage{tabularx, multirow}

\usepackage{siunitx}

\usepackage{amsmath,amssymb}
\usepackage[colorlinks=true,bookmarks=false,citecolor=blue,urlcolor=blue]{hyperref} 

\tikzstyle{block} = [rectangle, draw, fill=white, text width=12em, text centered, rounded corners, minimum height=2em]
\tikzstyle{arrow} = [thick,->,>=stealth]

\title{QoS-Aware Service Restoration in 5G Optical Transport Networks}

\author{
    \IEEEauthorblockN{\hspace*{-0.5cm} Zahra Sharifi Soltani$^\dagger$}
    \IEEEauthorblockA{\hspace*{-0.5cm} \textit{zahra\_sharifisoltani@student.uml.edu}}
    \and
    \IEEEauthorblockN{Arash Rezaee$^\dagger$}
    \IEEEauthorblockA{\textit{arash\_rezaee@student.uml.edu}}
    \and
    \IEEEauthorblockN{Orlando Arias$^\dagger$}
    \IEEEauthorblockA{\textit{orlando\_arias@uml.edu}}
    \and
    \IEEEauthorblockN{Vinod M. Vokkarane$^\dagger$}
    \IEEEauthorblockA{\textit{vinod\_vokkarane@uml.edu}} 
    \and
    \IEEEauthorblockA{\hspace{1.5cm}\centering \textit{$^\dagger$Electrical and Computer Engineering Department, University of Massachusetts Lowell, United States}
    }
}

\vspace{-1.cm}

\begin{document}

\maketitle
\renewcommand\thefootnote{}\footnotetext{This paper was partially supported by NSF project award \#2008530.}

\begin{abstract}
The rapid growth of high-bandwidth applications in fifth-generation (5G) networks and beyond has driven a substantial increase in traffic within transport optical networks. While network slicing effectively addresses diverse quality of service (QoS) requirements—including bit rate, latency, and reliability—it also amplifies vulnerabilities to failures, particularly when a single disruption in the optical layer impacts multiple services within the 5G network. To address these challenges, we propose a Fast Disrupted Service Prioritization (FDSP) algorithm that strategically allocates resources to the most critical disrupted services. Specifically, FDSP employs a fast-solving integer linear programming (ILP) model to evaluate three key factors—service priority, bit rate, and remaining holding time—and integrates a physical-layer impairment (PLI)-aware routing and spectrum allocation approach. By leveraging this combined strategy, FDSP minimizes service disruption while optimizing resource utilization. Simulation results on Germany’s network demonstrate that our approach significantly enhances the reliability and efficiency of survivable 5G slicing, thereby reducing blocking probability.
\end{abstract}

\section{Introduction}
The significant growth of data-driven services and the increasing use of bandwidth-demanding applications have led to a substantial rise in traffic within fifth-generation (5G) networks and beyond. To address this growing demand, elastic optical networks (EONs) has become the backbone of 5G core networks, enabling a wide range of critical applications, including ultra-reliable low-latency communications (URLLC), enhanced mobile broadband (eMBB), and massive machine-type communications (mMTC) \cite{ofc2019,gu2023}. Each of these applications has demanding performance requirements in terms of bandwidth, latency, and reliability, challenging the network's ability to allocate resources efficiently. To meet these requirements, the success of 5G core networks depends on flexible resource allocation through network slicing, which allows service providers to meet the diverse demands of various use cases while guaranteeing quality of service (QoS).

5G core networks face increasing traffic demands, but their expansion also amplifies vulnerabilities to failures. Addressing these challenges requires advanced recovery strategies, particularly in optical transport networks. Failures in optical networks can lead to cascading effects, where a single failure in the optical layer results in multiple failures across 5G network slices, simultaneously disrupting numerous services or use cases \cite{gangopadhyay2019}. Addressing these challenges requires advanced strategies to ensure both resilience and efficiency in 5G optical transport networks. In the case of URLLC, network failures may result in catastrophic outcomes, particularly for safety-critical applications such as autonomous vehicles and remote surgery. These potential disruptions highlight the urgent need for reliable and priority-based network recovery mechanisms.

To address these vulnerabilities, recent researches has focused on developing advanced strategies for network resilience and efficient recovery in optical transport networks. In \cite{tnsm-2020}, Shahriar et al. propose a bandwidth-squeezing and multi-path provisioning method, ensuring significant bandwidth recovery during failures while using only a minimal amount of additional spectrum. Amjad et al. introduce a spectrum-aware path restoration mechanism, enabling efficient traffic recovery in sixth-generation (6G) networks by utilizing spectrum redistribution and regeneration techniques in \cite{ondm-2023}. 
Dedicated virtual network topology protection was explored in \cite{jocn-2015}, offering fast recovery by reserving disjoint paths for protection, though at the cost of higher resource consumption.
Furthermore, Gangopadhyay et al. proposed a static network design approach for a hyper-scale architecture using high-density transponders, grooming algorithms, and a shared pool of 3R regenerators for optical restoration, significantly reducing infrastructure costs and power consumption while maintaining resilience against multiple failures in 5G transport networks \cite{gangopadhyay2019}.

Despite significant advancements, optimizing resource usage in optical networks during recovery, particularly based on service type prioritization, remains a major challenge and has not been thoroughly explored. We proposed the fast disrupted service prioritization (FDSP) algorithm to prioritize critical disrupted services based on priority, bit rate, and remaining holding time using integer linear programming (ILP), integrated with a heuristic PLI-aware routing and spectrum allocation (RSA) strategy. This approach enhances the reliability and resource utilization efficiency of service restoration in 5G optical transport networks. Additionally, the first detect first serve (FDFS) approach is used as a benchmark, where services are served in the order they are detected as disrupted.\vspace{-0.2cm}

\section{Proposed Algorithm}
In this section, we first present the FDSP algorithm, followed by the PLI-aware RSA approach. In the FDSP algorithm, we formulate the prioritization of disrupted services as a weight optimization problem using ILP, aiming to efficiently rank disrupted services for recovery based on three normalized variables: service priority ($p_{i}$), service bit rate ($b_{i}$), and remaining holding time ($t_{i}$). The objective function maximizes the combined weight of these factors, expressed as\vspace{-0.2cm}
\begin{equation}\label{eq:obj} 
    \text{Maximize} \quad  \sum_{i} (w_b b_{i} + w_t t_i + w_p p_i),\vspace{-0.1cm}
\end{equation}
where $w_b$, $w_t$, and $w_p$ represent the respective weights assigned to the bit rate, remaining holding time, and priority of disrupted services. These weights are subject to the following constraints:
\vspace{-1cm}
\begin{center}
\begin{align}
    w_b + w_t + w_p = 1 \label{eq:obj1}\\ 
    w_p \geq  w_b + w_t \label{eq:obj2} \\
    w_b \geq  w_t \label{eq:obj3}\\
    w_b, w_t, w_p \geq 0  \vspace{-0.2cm} \label{eq:obj4} 
\end{align}
\end{center}

The first constraint (Eq. \ref{eq:obj1}) ensures that the total weight distribution across the three variables—bit rate, remaining holding time, and priority—sums to 1, fully utilizing the available weight capacity. The second constraint (Eq. \ref{eq:obj2}) prioritizes the weight assigned to priority ($w_p$), ensuring it is at least as large as the combined weights of service bit rate ($w_b$) and holding time ($w_t$), reflecting the importance of prioritization. The third constraint (Eq. \ref{eq:obj3}) ensures that the weight assigned to bit rate is greater than or equal to that of holding time, highlighting service bit rate's relative importance. Lastly, the non-negativity constraint (Eq. \ref{eq:obj4}) ensures that all weights are positive, maintaining feasibility within the optimization model.
This ILP model ensures optimal resource allocation by assigning higher importance to services based on their priority while balancing requested bit rate and remaining holding time. The constraints reflect the relative importance of each variable in the recovery process. Once the ILP is solved, the optimal values of the weights are calculated. The algorithm then computes the weighted sum for each service by multiplying the optimal weights with the corresponding service variables. Services are then sorted in descending order based on their weighted sum, ensuring that the most critical services are prioritized for recovery.

The proposed PLI-aware algorithm is designed to allocate resources in EONs by selecting suitable paths and efficiently allocating spectrum resources, even in the event of network disruptions. This method optimizes resource allocation by minimizing path length (and thus delay) while leveraging pre-calculated transmission reach and slot requirements. The algorithm combines K-shortest paths (KSPs) strategy with a first-fit approach to allocate spectrum for incoming services. Initially, it identifies a set of KSPs between source and destination nodes. In case of a failure, the algorithm removes disrupted links from the topology and uses the remaining network to find alternative paths. For each path, it calculates the length and selects the optimal modulation format, along with the required number of spectrum slots for each bandwidth, using pre-calculated values for maximum optical transmission reach (OTR) and slot requirements for different modulation formats. The pre-calculated OTR values, adopted from \cite{lanman-2024}, are presented in Table \ref{table:reach}. This ensures that the allocation decisions account for both the required capacity and signal-to-noise ratio (SNR) while considering PLI constraints.

\begin{table}[t]
\caption{Optical transmission reach (OTR) (in \si{\kilo\meter}) \cite{lanman-2024}}
\centering
\begin{tabular}{*{4}{c}}
\toprule
\multirow{2}{*}{\textbf{Bit Rate}}
& \multicolumn{3}{c}{\textbf{Modulation Format}} \\
& \textbf{PM-QPSK} & \textbf{PM-16QAM} & \textbf{PM-64QAM} \\
\midrule
100 Gbps
& 5190 & 2324 & 876 \\
200 Gbps
& 2595 & 1162 & 438 \\
400 Gbps
& 1298 & 581 & 219 \\
\bottomrule
\end{tabular}
\label{table:reach}
\vspace{-0.5cm}
\end{table}

Once the paths are identified, the algorithm allocates the necessary spectrum resources. It checks each path in the candidate set to verify whether sufficient contiguous spectrum slots are available. The number of slots required is calculated using the highest feasible modulation format for the path and bandwidth requirements. A first-fit method is then applied to find suitable continuous and contiguous block. Upon finding such a block, the algorithm assigns the path and spectrum slots, returning the path and the calculated slot indices. If no suitable path and spectrum is available, the service is blocked.

\begin{figure*}[tp] 
    \centering
    \begin{subfigure}[t]{0.24\textwidth}
        \centering
        \begin{tikzpicture} [scale=0.7]
        \begin{axis}[
         xlabel={Network Load [Erlang] },
         ylabel={Restoration BBP},
         ylabel style={yshift=-2.5ex},
         xmin=50, xmax=1000,
         ymin= , ymax= ,
         xtick={},
         x tick label style={align=center,text width= 0.5cm},
         ytick={},
         legend pos=north west,
         legend style={at={(0.65,0.15)},anchor=west},
         ymajorgrids=true,
         xmajorgrids=true,
         grid style=dotted,
         title style={at={(0.5,-0.22)},anchor=north,yshift=-0.1},
        table/col sep=comma,
        height=2.45 in,
        ]
        \addplot [red,mark=o,] table[y=bb_p3] {results-256/blocking/benchmark_4_blocking.csv};
        \addplot [blue,mark=o,] table[y=bb_p3] {results-256/blocking/ILP_4_blocking.csv};
        \legend{{\small FDFS}, \small FDSP,}
        \end{axis}
        \end{tikzpicture}
        \caption{Four failures/Priority 3.}
        \label{fig:f4p3}
    \end{subfigure}
    \hfill
    \begin{subfigure}[t]{0.24\textwidth}
    \centering
        \begin{tikzpicture}[scale=0.7]
        \begin{axis}[
         xlabel={\normalsize Network Load [Erlang] },
         ylabel={~},
         ylabel style={yshift=-2.5ex},
         xmin=50, xmax=1000,
         ymin= , ymax= ,
         xtick={},
         x tick label style={align=center,text width= 0.5cm},
         ytick={},
         legend pos=north west,
         legend style={at={(0.65,0.15)},anchor=west},
         ymajorgrids=true,
         xmajorgrids=true,
         grid style=dotted,
         title style={at={(0.5,-0.22)},anchor=north,yshift=-0.1},
        table/col sep=comma,
        height=2.45in,
        ]
        \addplot [red,mark=o,] table[y=bb_p3] {results-256/blocking/benchmark_3_blocking.csv};
        \addplot [blue,mark=o,] table[y=bb_p3] {results-256/blocking/ILP_3_blocking.csv};
        \legend{{\small FDFS}, \small FDSP,}
        \end{axis}
        \end{tikzpicture}
        \caption{Three failures/Priority 3.}
        \label{fig:3fp3}
    \end{subfigure}
    \hfill
    \begin{subfigure}[t]{0.24\textwidth}
        \centering
        \begin{tikzpicture}[scale=0.7]
        \begin{axis}[
         xlabel={\normalsize Network Load [Erlang] },
         ylabel={},
         ylabel style={yshift=-2.5ex},
         xmin=50, xmax=1000,
         ymin= , ymax= ,
         xtick={},
         x tick label style={align=center,text width= 0.5cm},
         ytick={},
         legend pos=north west,
         legend style={at={(0.65,0.15)},anchor=west},
         ymajorgrids=true,
         xmajorgrids=true,
         grid style=dotted,
         title style={at={(0.5,-0.22)},anchor=north,yshift=-0.1},
        table/col sep=comma,
        height=2.45in
        ]
        \addplot [red,mark=o,] table[y=bb_p1] {results-256/blocking/benchmark_4_blocking.csv};
        \addplot [blue,mark=o,] table[y=bb_p1] {results-256/blocking/ILP_4_blocking.csv};
        \legend{{\small FDFS}, \small FDSP,}
        \end{axis}
        \end{tikzpicture}
        \caption{Four failures/Priority 1.}
        \label{fig:graph3}
    \end{subfigure}
    \hfill
    \begin{subfigure}[t]{0.24\textwidth}
        \centering
        \begin{tikzpicture}[scale=0.7]
        \begin{axis}[
         xlabel={\normalsize Network Load [Erlang] },
         ylabel={},
         ylabel style={yshift=-2.5ex},
         xmin=50, xmax=1000,
         ymin= , ymax= ,
         xtick={},
         x tick label style={align=center,text width= 0.5cm},
         ytick={},
         legend pos=north west,
         legend style={at={(0.65,0.15)},anchor=west},
         ymajorgrids=true,
         xmajorgrids=true,
         grid style=dotted,
         title style={at={(0.5,-0.22)},anchor=north,yshift=-0.1},
        table/col sep=comma,
        height=2.45in,
        ]
        \addplot [red,mark=o,] table[y=bb_p1] {results-256/blocking/benchmark_3_blocking.csv};
        \addplot [blue,mark=o,] table[y=bb_p1] {results-256/blocking/ILP_3_blocking.csv};
        \legend{{\small FDFS}, \small FDSP,}
        \end{axis}
        \end{tikzpicture}
        \caption{Three failures/Priority 1.}
        \label{fig:graph4}
    \end{subfigure}
    \caption{Restoration bit rate blocking probability for 3 and 4 failures and priorities 1 and 3.\vspace{-0.2cm}}
    \label{fig:all_graphs}

\end{figure*}
\vspace{-0.2cm}
\begin{table*}[htbp]
    \centering
    \caption{Percentage change in recovered holding time of FDSP vs. FDFS for 4 and 3 failures, URLLC and mMTC.}
{\small
\setlength{\tabcolsep}{4.5pt}
    \begin{subtable}[t]{0.565\linewidth}\vspace{-1ex}
        \centering
        \caption{Four Failures in Germany 50 Topology\vspace{-0.2cm}}
        \label{table:ht_a}
        \begin{tabularx}{\columnwidth}{@{}p{0.13\linewidth}*{6}{c}@{}}
            \toprule
            \multicolumn{1}{c}{}&\multicolumn{3}{c}{Priority 3 (URLLC)} & \multicolumn{3}{c}{Priority 1 (mMTC)} \\ 
            &Max&Min&Average&Max&Min&Average\\
            \midrule
            RHT (\%)&10&0.24&4.66&-10.39&-0.63&-5.16 \\
            Erlang&1000&150&[50, 1000]&950&100&[50, 1000]\\
            \bottomrule
        \end{tabularx}
    \end{subtable}\hfill%
    \begin{subtable}[t]{0.429\linewidth}\vspace{-1ex}
        \centering
        \caption{Three Failures in Germany 50 Topology\vspace{-0.2cm}}
        \label{table:ht_b}
        \begin{tabularx}{\columnwidth}{@{}*{6}{c}@{}}
            \toprule
            \multicolumn{3}{c}{Priority 3 (URLLC)} & \multicolumn{3}{c}{Priority 1 (mMTC)} \\  
            Max&Min&Average&Max&Min&Average\\
            \midrule
            8.99&0.41&4.046&-9.51&-0.062&-4.6\\
            950&100&[50, 1000]&1000&100&[50, 1000]\\
            \bottomrule
        \end{tabularx}
    \end{subtable}
    }

    \label{table:ht}
    \vspace{-0.35cm}
\end{table*}

\section{Simulation Results}\vspace{-0.2cm}
The presented algorithm is evaluated using extensive numerical simulations in OpticalRL-gym \cite{optrlgym-2020} on the Germany network topology with 50 nodes, with fiber links equipped with 256 frequency slots, each
\SI{12.5}{\giga\hertz}
bandwidth. The mean holding time is
\SI{3600}{\second}, and services have bandwidths of 100, 200, or 400 $Gbps$, following a uniform distribution. Priorities are assigned as mMTC (1), eMBB (2), and URLLC (3), with a ratio of 25:40:35 \cite{thantharate2019}. Restoration bit rate blocking probability (BBP) is measured over 5000 independent iterations, each handling 5000 requests 
network failures and occur after 3000 requests when the network reaches a steady state. BBP is calculated by dividing the bit rate of restored services by the bit rate of disrupted services. Modulation formats include PM-QPSK, PM-16QAM, and PM-64QAM, with spectral efficiencies of 2, 3, and 6 per polarization, and one slot is used as a guard band. Before analyzing the results, it is worth noting that simulation results confirm the algorithm's speed, efficiently processing more than 70 requests within \SI{3}{\milli\second}.
    
Fig. \ref{fig:f4p3} and \ref{fig:3fp3} show the restoration BBP versus traffic load for priority 3 disrupted demands under 4 and 3 independent failures in the Germany topology, while Fig.
\ref{fig:graph3} and \ref{fig:graph4}
present the same for priority 1 disrupted demands comparing the FDSP and FDFS approaches. Fig. 
\ref{fig:f4p3} and \ref{fig:3fp3}
demonstrate that the FDSP algorithm, by prioritizing services with priority 3 and higher bit rates to be served earlier compared to other lower-priority services, achieves a lower BBP compared to the FDFS approach. FDSP consistently outperforms the benchmark (FDFS) across all traffic levels, with a particularly significant advantage at higher loads. For example, at 1000 Erlangs, FDSP reduces the restoration BBP by approximately 16\% in the case of four failures and by 14\% with three failures. On average, the restoration BBP for disrupted URLLC services (priority 3) is reduced by 12\% across all traffic loads ranging from 50 to 1000 Erlangs in the network with 4 failures, compared to the FDFS approach. In the case of 3 failures, the restoration BBP for URLLC services is lowered by approximately 10\% over the same traffic load range. 

Due to space limitations, we have excluded the restoration BBP results for disrupted eMBB services (priority 2). However, it's important to note that the BBP for these services remains consistent compared to the benchmark. The simulation results in Fig.
\ref{fig:graph3} and \ref{fig:graph4}
show that the restoration BBP for disrupted mMTC services (lowest priority) increases when utilizing the FDSP approach compared to the benchmark, in both the 3 and 4 failure scenarios. On average, across all traffic loads from 50 to 1000 Erlangs, the restoration BBP for mMTC services increases by approximately 10\% and 9\% in the 4 and 3 failure scenarios, respectively. The simulation results show that FDSP efficiently allocates resources during failures, prioritizing high-priority, high-bandwidth services.

Table \ref{table:ht} summarizes the improvement ratio of recovered service's remaining holding time (RHT) for disrupted priority 3 and 1 services (URLLC and mMTC) in the FDSP approach compared to the benchmark for 4 and 3 failures (Tables \ref{table:ht_a} and \ref{table:ht_b}). The RHT ratio is defined as the remaining holding time of recovered services divided by the remaining holding time of disrupted services. Remaining holding time is the third criterion used to prioritize disrupted services, and due to space limitations, the results are presented in the table. The largest improvement is observed at 1000 and 950 Erlangs, with increases of 10\% and 8.87\% over the benchmark for 4 and 3 failures, respectively. On average, across all loads from 50 to 1000 Erlangs, RHT improved by 4.66\% and 4.04\% compared to FDFS for 4 and 3 failures. The algorithm performs better under high loads due to the scarcity of available resources, while in lower loads, the improvement is modest, at less than 0.5\%. As expected from the observation of blockings, RHT ratio for priority 1 (mMTC) services decreases in FDSP compared to the benchmark, particularly at high loads, as detailed in Table \ref{table:ht} for both scenarios. The largest reduction is observed at 1000 and 950 Erlangs, with decreases of approximately 10.39\% and 9.51\% for the 4 and 3 failure scenarios, respectively. On average, across all loads from 50 to 1000 Erlangs, RHT ratio decreased by 5.16\% and 4.6\% for the 4 and 3 failure scenarios. However, the average recovered holding time for priority two remains the same as the benchmark.

\vspace{-0.0cm}
\section{Conclusion}
\vspace{-0.0cm}
We presented the FDSP algorithm to prioritize the disrupted services for efficient restoration in 5G optical transport networks. Simulations show that FDSP consistently outperforms FDFS by ranking disrupted services based on priority, bit rate, and remaining holding time using ILP, solving over 70 services in less than \SI{3}{\milli\second}. In the Germany topology with four failures, BBP decreases by 16\% at 1000 Erlang. Finally, future work can explore scalability for 6G use cases, diverse traffic priorities, and additional topologies.\vspace{-0.1cm}

\bibliographystyle{IEEEtran}
\bibliography{refrence}
{

\end{document}